\documentclass[twocolumn,showpacs,aps,superscriptaddress,pra,floatfix,noshowpacs,showkeys]{revtex4}
\usepackage{epsfig}
\usepackage{amsmath}
\usepackage{amssymb}
\usepackage{bm}
\usepackage{graphicx}

\begin{document}

\title{Simulation of the relativistic electron dynamics and acceleration in a linearly-chirped laser pulse}

\author{Najeh M. Jisrawi}
\affiliation{Department of Applied Physics, University of Sharjah, POB 27272, Sharjah, United Arab Emirates}

\author{Benjamin J. Galow}
\affiliation{Gaisbergstra{\ss}e 61, 69115 Heidelberg, Germany}

\author{Yousef I. Salamin}
\affiliation{Department of Physics, American University of Sharjah, POB 26666, Sharjah,
United Arab Emirates}
\date{\today}

\begin{abstract}

Theoretical investigations are presented, and their results are discussed, of the laser acceleration of a single electron by a chirped pulse. Fields of the pulse are modeled by simple plane-wave oscillations and a $\cos^2$ envelope. The dynamics emerge from analytic and numerical solutions to the relativistic Lorentz-Newton equations of motion of the electron in the fields of the pulse. All simulations have been carried out by independent Mathematica and Python codes, with identical results. Configurations of acceleration from a position of rest as well as from injection, axially and sideways, at initial relativistic speeds are studied. \\

\end{abstract}


\maketitle

\section{Introduction}\label{sec:intro}

Particle laser-acceleration is currently an active field of research, stimulated by the continued progress made in laser technology (Yanovsky {\it et al.}, 2008) and the need for machines that are more compact and less costly than the conventional accelerators. Many schemes and a number of different accelerator configurations have been suggested and investigated both theoretically (Scully \& Zubairy, 1991; Umstadter {\it et al.}, 1995; Esarey {\it et al.}, 1995; Salamin {\it et al.}, 2000; Wang {\it et al.}, 2001; Pang {\it et al.}, 2002; Salamin \& Keitel, 2002; Kawata {\it et al.}, 2005; Salamin, 2006; Xu {\it et al.}, 2007; Salamin {\it et al.}, 2008; Xie {\it et al.}, 2010) and experimentally (Malka {\it et al.}, 1997; Shao {\it et al.}, 2013; Plettner {\it et al.}, 2005; Sears {\it et al.}, 2005). The idea of employing a chirped pulse in laser acceleration has been suggested relatively recently (Singh, 2005; Sohbatzadeh {\it et al.}, 2006; Sohbatzadeh {\it et al.}, 2009; Gupta \& Suk, 2007; Sohbatzadeh {\it et al.}, 2010; Sohbatzadeh {\it et al.}, 2011; Galow {\it et al.}, 2011; Salamin, 2012). The basic mechanism at work in this scheme hinges on the fact that chirping the frequency distorts the pulse and generates a quasi-static electric field portion which, in turn, gives the particle, e.g., an electron, a substantial boost. In other words, the electron moves synchronously with the pulse by {\it surfing} over this low-frequency portion and gains energy continuously from it. The scheme, as such, works to accelerate electrons from rest, and can be useful as a booster, too.

Several papers have been written on acceleration by a chirped pulse, mostly presenting numerical results. In this paper, the aim is two-fold. Efforts will be invested here to find out how much progress can be achieved by pursuing the analytic investigations. It is always good to develop the equations that guide our thinking about the physics and also help us benchmark the numerical simulations, which will ultimately be resorted to, to handle a real accelerator configuration. The second, intimately related, aim of this work is to some extent pedagogical. It is hoped that detailed investigation of the various dynamical aspects of the process will shed new and much needed light onto the mechanism of acceleration by a chirped pulse (Galow {\it et al.}, 2011).

The basic theory and main working equations will be presented in Sec. \ref{sec:basic}. In the same section, the equations will be used to investigate the dynamics of a single electron injected axially for subsequent interaction with a chirped pulse. The parameters used in this investigation are the same as those used elsewhere (Hartemann {\it et al.}, 1995) albeit employing a non-chirped pulse. The purpose is to show that both investigations agree in the appropriate limits. Details of the investigation of acceleration configurations of electrons initially at rest at the origin of coordinates, axially injected, and injected sideways at some angle with respect to the propagation direction of the pulse, will be presented in Sec. \ref{sec:dynamics}. Finally, a summary and the main conclusions will be given in Sec. \ref{sec:discussion}.

\section{Basic Theory}\label{sec:basic}

The theoretical background of laser acceleration by a chirped plane-wave laser pulse will be briefly described in this section. A particular case of axial injection will be discussed in some detail and the results of our calculations will be shown to have the correct limits of acceleration by the corresponding unchirped pulse (Hartemann {\it et al.}, 1995).

\subsection{General}\label{sec:general}

In the main working equations for the dynamics of a single electron in chirped plane-wave laser fields, the electron is treated as a point particle of mass $m$ and charge $-e$, the relativistic energy and momentum are given, respectively, by ${\cal E}=\gamma mc^2$ and $\bm{p}=\gamma mc\bm{\beta}$, where $\bm{\beta}$ is the velocity of the particle scaled by $c$, the speed of light in vacuum, and $\gamma=(1-\beta^2)^{-1/2}$. The fields, on the other hand, are modeled by an infinite plane-wave and a finite-duration plane-wave pulse. In modeling those fields the combination $\eta=\omega_0t-k_0z$, in which $\omega_0$ is the initial (unchirped) frequency and $k_0=2\pi/\lambda_0$ is the wavenumber, is used as a variable. Thus, the chirped frequency is $\omega=\omega_0(1+b\eta)$, where $b$ is the dimensionless chirp parameter. We work with the fields (SI units)

\begin{eqnarray}
 \label{E} \bm{E}(\eta) &=& \hat{\bm{i}} E_0\sin(\eta+b\eta^2)
    \cos^2\left[\frac{\pi}{\tau\omega_0}(\eta-\bar{\eta})\right], \\
 \label{B} \bm{B}(\eta) &=& \hat{\bm{j}} \frac{E(\eta)}{c}.
\end{eqnarray}
These equations model the fields of a pulse with a $\cos^2$ envelope, a temporal width $\tau = 50$ fs, a wavelength $\lambda_0 = 1~\mu$m, provided the choice $\bar{\eta} = 15\pi$ is made. For an electron interacting with an infinite plane wave, the normalized field strength $a=eE_0/mc\omega_0$ is related to the laser field intensity $I$ by $a=e\sqrt{2I/c\varepsilon_0}/(mc\omega_0)$, where $\varepsilon_0$ is the permittivity of free space and $\lambda_0$ is the wavelength of the unchirped laser field. Inserting values of the universal constants, one gets
\begin{equation}\label{I}
    I\left[\frac{\text{W}}{\text{cm}^2}\right] = 1.36817\times10^{18}\left[\frac{a}{\lambda_0[\text{$\mu$m}]}\right]^2,
\end{equation}
thus making $a^2$ a dimensionless intensity parameter.

\begin{figure}[t]
\includegraphics[width=8cm]{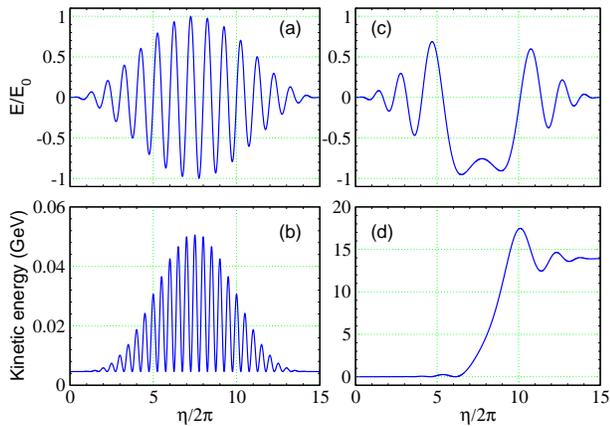}
\caption{(Color online) (a) Evolution of the normalized unchirped ($b = 0$) electric field of a $\cos^2$ laser pulse of duration $\tau=50$ fs. (b) Evolution of the electron kinetic energy during interaction with the unchirped pulse. (c) and (d): Same as (a) and (b), respectively, but for a chirped laser pulse ($b = -0.0103$). The remaining parameters are: $\lambda_0=1~\mu$m, $a=3$ (intensity $I\sim1.23\times10^{19}$ W/cm$^2$) and $\gamma_0=10$ (injection kinetic energy $K_0\sim4.6$ MeV). }
\label{fig1}
\end{figure}

In the next two subsections, two sets of conditions on the initial injection position and velocity of the electron will be considered. Sec. \ref{sec:axial} will be devoted to the axial injection configuration, while the more general case of electron injection at an angle $\zeta_0$ to the pulse propagation direction, will be taken up in Sec. \ref{sec:sideways}. Various aspects of the electron dynamics will be discussed, including evolution of its kinetic energy and velocity components, as well as its trajectories. In order to be able to tackle the general injection situation, for which some of the analytic solutions become too cumbersome, we will resort here to numerical integration of the combined equations of motion
\begin{equation}\label{betaeq}
    \frac{d\bm{\beta}}{dt}=\frac{e}{\gamma mc}\left[\bm{\beta}(\bm{\beta}\cdot\bm{E})-(\bm{E}+c\bm{\beta}\times\bm{B})\right].
\end{equation}
Instead of working directly with the time $t$ as a variable, we will employ $\eta$. The integration limits will be denoted by $\eta_i$ and $\eta_f$. Equation (\ref{betaeq}) is equivalent to three coupled
differential equations. This system of differential equations has been solved numerically using Runge-Kutta-based Mathematica codes we have developed over the years. All of our results have independently been confirmed by codes written in Python, as well.

\subsection{Axial injection}\label{sec:axial}

In (Salamin, 2012) the special case corresponding to $\zeta_0=0$ (axial injection) was briefly discussed. Much of the discussion there was based on the analytic solution to the Newton-Lorentz equations. Issues pertaining to this particular case not covered in (Salamin, 2012) will be discussed here for completeness and in order to demonstrate that the results obtained in the chirped pulse have their correct unchirped limits published elsewhere (Hartemann {\it et al.}, 1995).

\begin{figure}[t]
\includegraphics[width=8cm]{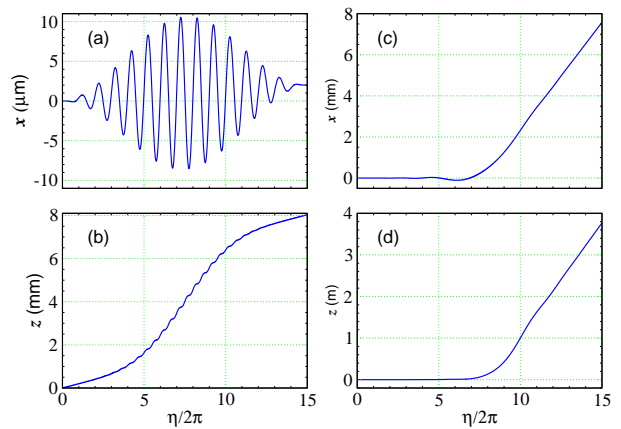}
\caption{(Color online) (a) and (b): Evolution of the transverse and axial position coordinates, respectively, as functions of the variable $\eta$, while interacting with an unchirped ($b = 0$) $\cos^2$ laser pulse of duration $\tau = 50$ fs. (c) and (d): Same as (a) and (b), respectively, but for a chirped pulse ($b = -0.0103$). The remaining parameters are the same as in Fig. \ref{fig1}.}
\label{fig2}
\end{figure}

For the initial conditions on particle injection the assumption will be made that the front of the pulse catches up with the MeV electron at $t=0$, precisely at the origin of coordinates ($x_0=y_0=z_0=0$). Hence, $\eta_i=0$ and $\eta_f=30\pi$ for the $\cos^2$ pulse described above. furthermore, the electron is assumed to be traveling axially (along the direction of propagation of the pulse, the $z-$axis) with a scaled speed $\beta_0$, derived from $\gamma_0=10$, which corresponds to the initial injection kinetic energy $K_0\sim4.6$ MeV. For an unchirped laser pulse of field intensity $I\sim1.23\times10^{19}$ W/cm$^2$ ($a=3$) this configuration has been discussed thoroughly elsewhere (Hartemann {\it et al.}, 1995). What we will do below is investigate the same electron dynamics, albeit in a chirped laser pulse. The results will be displayed along with their unchirped counterparts, so that visual comparison may be facilitated.

\begin{figure}
\vskip0.3cm
\includegraphics[width=8cm]{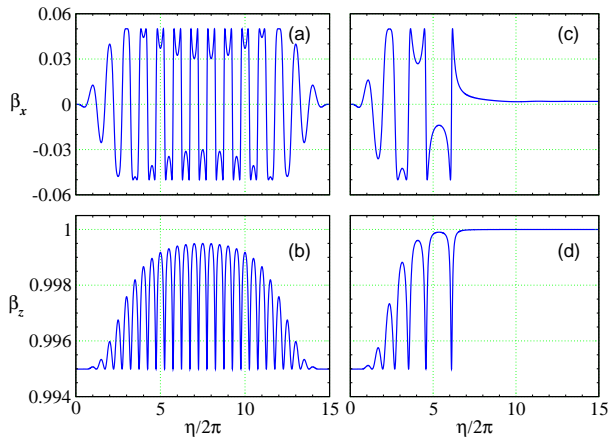}
\caption{(Color online) (a) and (b): Evolution of the electron transverse and axial scaled velocity components, respectively, as functions of the variable $\eta$, while interacting with an unchirped ($b = 0$) $\cos^2$ laser pulse of duration $\tau = 50$ fs. (c) and (d): Same as (a) and (b), respectively, but for a chirped pulse ($b = -0.0103$). The remaining parameters are the same as in Fig. \ref{fig1}.}
\label{fig3}
\end{figure}

To begin with, Fig. \ref{fig1}(a) shows the unchirped scaled electric field of the pulse vs. $\eta$. Fig. \ref{fig1}(c) shows how this field gets distorted when the frequency is chirped using $b=-0.0103$. The pulse develops an asymmetrical low-frequency, indeed quasi-static, part. In Fig. \ref{fig1}(b) evolution with $\eta$ of the kinetic energy of the electron is shown for $b=0$. As expected from the Lawson-Woodward theorem (Woodward, 1947; Lawson, 1979) the electron gains no net energy from interaction with the plane-wave unchirped pulse. Substantial gain, however, is shown in Fig. \ref{fig1}(d) to result from synchronous interaction with the low-frequency part of the chirped pulse. As expected, too, interaction with the wings, whose frequency is minimally affected by the chirp, results in very little or no further energy gain at all.

As has been shown elsewhere (Hartemann {\it et al.}, 1995; Salamin, 2012), in the case of axial injection, the trajectory of the electron is confined to the $xz-$plane, the polarization plane of the laser pulse. Instead of showing the $xz$ trajectory, Fig. \ref{fig2} displays evolution of the $x-$ and $z-$ coordinates, separately, with $\eta$. By comparing the unchirped Fig. \ref{fig2}(a) with the chirped Fig. \ref{fig2}(c) one can read clearly the amount of net electron displacement, $\Delta x$, as a result of interaction with the pulse (poderomotive scattering). For the parameter set used, $\Delta x\sim2~\mu$m (unchirped) whereas $\Delta x\sim7.5~$mm (chirped). Likewise, the corresponding axial excursions made by the electron during interaction with the pulse are: $\Delta z\sim8~$mm (unchirped) and $\Delta z\sim3.8~$m (chirped).

\begin{figure}[t]
\includegraphics[width=8cm]{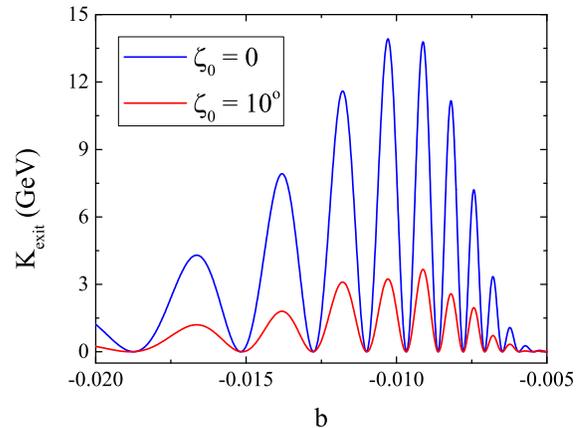}
\caption{(Color online) Electron exit kinetic energy, $K_{exit}\equiv K(\eta_f)=[\gamma(\eta_f)-1]mc^2$, where $\eta_f=30\pi$, vs. the chirp parameter, as a result of interaction with a chirped plane-wave laser pulse of a $\cos^2$ envelope and duration $\tau = 50$ fs. The laser wavelength is $\lambda_0=1~\mu$m, and $a=3$ ($I\sim1.23\times10^{19}$ W/cm$^2$). Initial injection is at $\gamma_0=10$ ($K_0\sim4.6$ MeV) axially ($\zeta_0=0$) and sideways at $\zeta_0=10^\circ$. Onset of the interaction is assumed to have been at $t=0$, the instant the electron passes through the origin of coordinates.}
\label{fig4}
\end{figure}

Components of the electron scaled velocity vector are displayed in Fig. \ref{fig3}. Interaction with the unchirped pulse gives results identical to those obtained earlier (Hartemann {\it et al.}, 1995). Figs. \ref{fig3}(a) and (b) show that the electron is left behind the pulse moving at exactly the same velocity with which it was injected initially, again as expected from the Lawson-Woodward theorem (Woodward, 1947; Lawson, 1979). By contrast, interaction with the chirped pulse results in clear particle acceleration, as shown in Figs. \ref{fig3}(c) and (d). The electron emerges from the interaction with a slight increase in its transverse velocity $\beta_x$, while $\beta_z\to 1$.

\subsection{Sideways injection}\label{sec:sideways}

\begin{figure}[t]
\includegraphics[width=8cm]{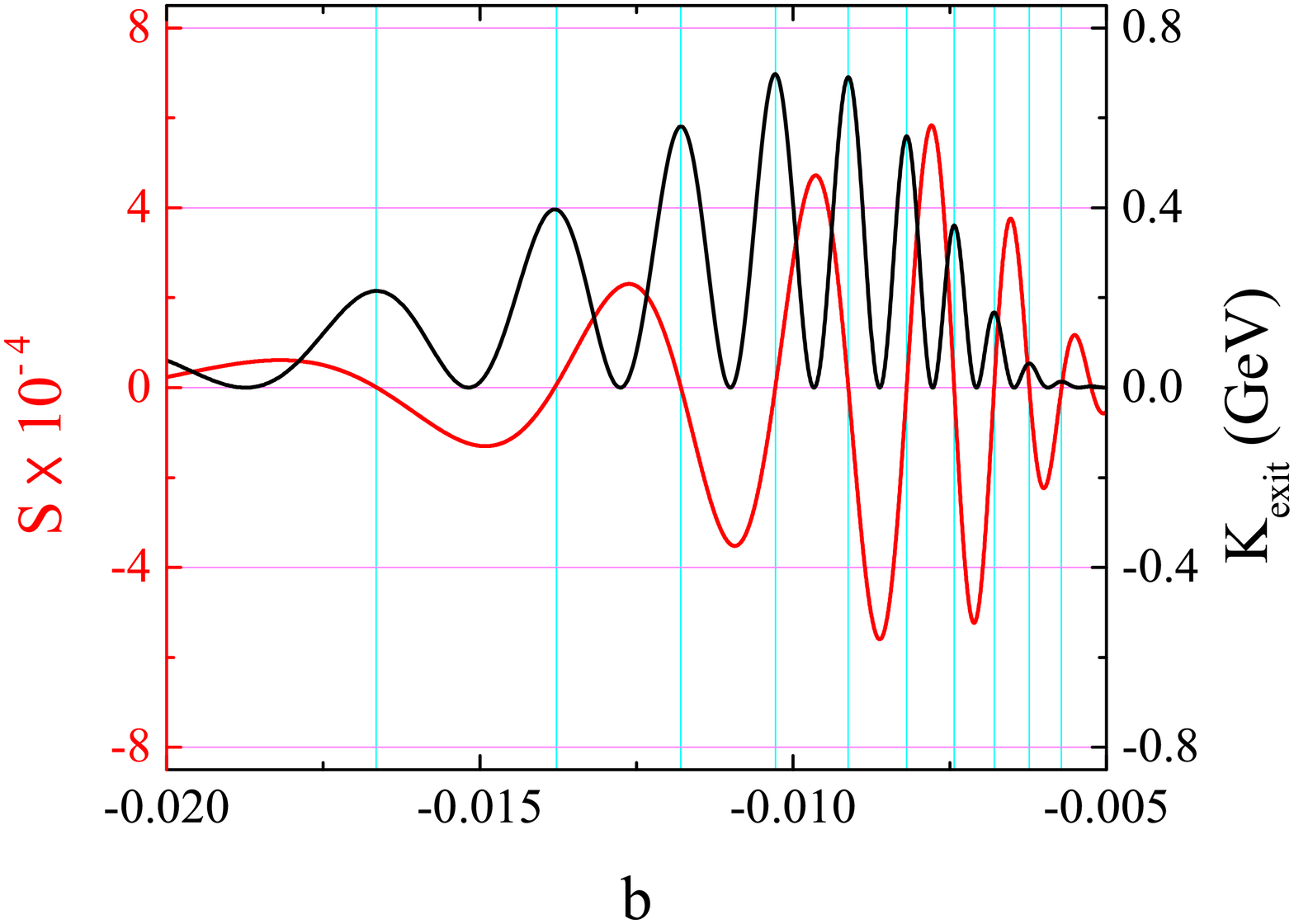}
\caption{(Color online) Same as Fig. \ref{fig4}, but for electron initial conditions of rest at the origin of coordinates. Shown also is the function $S(b)$ given by Eq. (\ref{S}) whose zeros are the values of $b$ corresponding to each of which the exit kinetic energy attained is a maximum. The vertical lines indicate the zeros of $S$ and the exit kinetic energy maxima.}
\label{fig5}
\end{figure}

The axial injection configuration, discussed in Sec. \ref{sec:axial}, is probably difficult to realize experimentally. Assuming that the model itself is sound, a more practically-realizable configuration would be one in which the electron is injected in, for example, the $yz-$plane at an angle $\zeta_0$ with the laser pulse propagation direction, the $z-$axis. Thus, the initial injection velocity of the electron, scaled by the speed of light $c$, may be written as

\begin{equation}\label{beta0}
  \bm{\beta}_0 = \beta_0(-\hat{\bm{j}}\sin\zeta_0+\hat{\bm{k}}\cos\zeta_0).
\end{equation}
Here, too, the assumption will be made that the front of the laser pulse catches up with the electron at $t=0$ at the origin of coordinates. Every result to be presented below has been produced employing numerical codes which return the corresponding result discussed in Sec. \ref{sec:axial} in the limit $\zeta_0\to0$. More importantly, the following logical amendments to the working equations must be introduced (Salamin, 2012). Adopting Eq. (\ref{beta0}) as the initial condition on the velocity requires that the constants of the motion be altered to read: $c_1=\gamma\beta_y=-\gamma_0\beta_0\sin\zeta_0$, and $c_2=\gamma(1-\beta_z)=\gamma_0(1-\beta_0\cos\zeta_0)$. Taking all of this into consideration, the scaled energy expression takes the form
\begin{equation}\label{gamma}
  \gamma=\frac{1+(\gamma\beta_x)^2+(\gamma_0\beta_0\sin\zeta_0)^2+
    \gamma_0^2(1-\beta_0\cos\zeta_0)^2}{2\gamma_0(1-\beta_0\cos\zeta_0)}.
\end{equation}
On the other hand, the general expression of $\gamma\beta_x(\eta)$ stays the same (Salamin, 2012)
\begin{equation}\label{gbx}
  \gamma\beta_x(\eta) = a\int_{\eta_0}^{\eta}\sin(\eta'+b\eta'^2)\cos^2\left[\frac{\pi}{\tau\omega_0}(\eta'-\bar{\eta})\right]d\eta'.
\end{equation}

Before investigating the electron dynamics in this general scenario, values of the chirp parameter $b$, appropriate for obtaining substantial acceleration, will have to be obtained. To this end, exit electron kinetic energies will be calculated as $b$ is varied from $-$0.04 to zero. The results are shown in Fig. \ref{fig4}, for the cases of axial injection ($\zeta_0=0^\circ$) and sideways injection (at $\zeta_0=10^\circ$). All other parameters are the same as in Sec. \ref{sec:axial}. The peaks in the figures occur at $b$ values which result in optimal synchronous motion between the electron and the pulse. Conversely, total lack of synchronization leads to the appearance of the minima, which express zero net energy gain. Note that the absolute maximum in Fig. \ref{fig4}, which has already been used in the case of axial injection, occurs at $b\sim-0.0103$.

\begin{figure}[t]
\includegraphics[width=8cm]{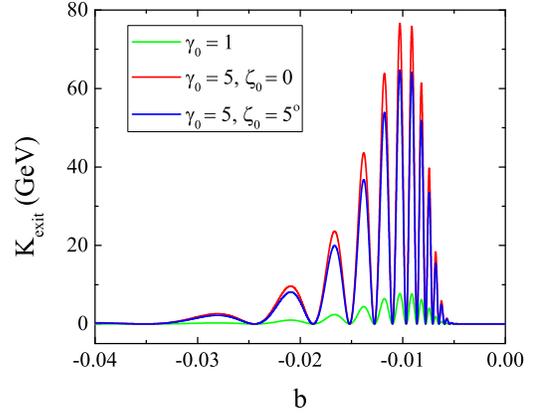}
\caption{(Color online) Same as Fig. \ref{fig4}, but for $a=10$ and electron initial conditions of rest at the origin of coordinates, and injection at $\gamma_0=5$, for $\zeta_0=0$ and $5^\circ$. }
\label{fig6}
\end{figure}

As it turns out, the plots of $K_{exit}$ vs. $b$, corresponding to the different cases of particle injection into the laser field, have the same structure and exhibit minima and maxima at precisely the same values of $b$, with all other parameters the same. The heights of the corresponding maxima, however, are different. For the chosen parameters, the exit kinetic energies associated with the configuration of sideways injection at $\zeta_0=10^\circ$ are roughly a factor of 4 smaller than those of the axial injection case.

The above conclusions should come as no surprise, in light of the following. To find the values of $b$ that correspond to maximum exit kinetic energies, one sets equal to zero the derivative with respect to $b$ of the expression of $\gamma$ given in Eq. (\ref{gamma}). This is tantamount to extremizing $\gamma\beta_x$ with respect to $b$, Eq. (\ref{gbx}). In fact, this process amounts to finding the zeros of the following function, obtained by differentiating with respect to $b$ under the integral in Eq. (\ref{gbx})
 \begin{equation}\label{S}
   S(b) = \int_{\eta_0}^{\eta}\eta'^2\cos(\eta'+b\eta'^2)\cos^2\left[\frac{\pi}{\tau\omega_0}(\eta'-\bar{\eta})\right]d\eta'.
 \end{equation}
 Careful inspection of $S(b)$ quickly reveals that it is independent of the initial velocity of the electron and the laser field intensity parameter $a$. Thus, those zeros as well as the maxima exhibited by the $K_{exit}$ vs. $b$ plots, will always have the same general structure and the same $b$ values at which the zeros of $S(b)$ and maxima of $K_{exit}$ occur, provided the pulse-shape is the same. In other words, the positions of the maxima are model-dependent. The heights of the maxima, however, depend upon $a^2$, $\gamma_0$ and $\zeta_0$. This is demonstrated in Fig. \ref{fig5} where, in addition to $S(b)$, the exit kinetic energy is shown over the same range of $b$ values for the case of an electron initially at rest at the origin. Vertical lines are added to indicate the zeros of $S(b)$ and the corresponding exit kinetic energy maxima.

\section{Further Electron Dynamics}\label{sec:dynamics}

In the investigations of the previous section the parameters used were chosen for the purpose of comparing our results, in the limit of $b\to0$ (unchirped configuration) with known ones (Hartemann, 1995), in addition to investigating the process of net energy gain from interaction with the chirped pulse. This section will be devoted to further investigations of the electron dynamics in chirped pulses employing a totally different, yet experimentally available, set of laser and electron parameters. The laser field intensity parameter will be taken as $a=10$ ($I\sim1.37\times10^{20}$ W/cm$^2$, at $\lambda_0=1~\mu$m).

Here, too, situations corresponding to three different sets of initial conditions will be explored. Variations of $K_{exit}$ with $b$ for all three cases are shown in Fig. \ref{fig6}. As has been shown above, the same values of $b$ locate the maxima of all cases. Axial injection results in best gains while less energy is gained when injection is made at $\zeta_0>0$. With increasing injection angle more energy is carried by the transverse degree of freedom, oscillation in the polarization direction of the laser pulse. Note also that the figure shows clearly that substantial energy is also gained in the case of initial rest at the origin. This configuration will be discussed in more details in the next subsection.

\begin{figure}
\includegraphics[width=8cm]{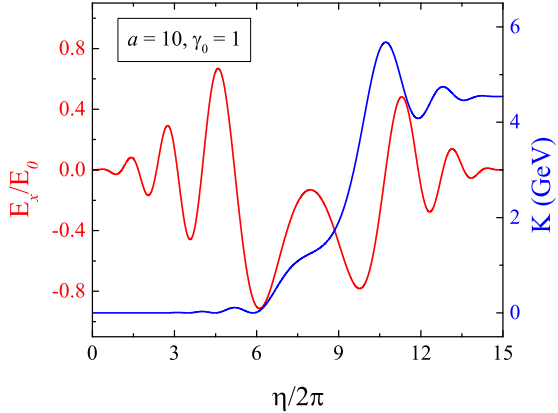}
\caption{(Color online) Evolution in $\eta$ of the normalized electric field strength, $E_x/E_0$, of a chirped ($b=-0.01$) $\cos^2$ laser pulse of 50 fs duration, and the kinetic energy, $K$, of a single electron interacting with it. The field intensity is $I\sim1.37\times10^{20}$ W/cm$^2$ ($a=10, \lambda_0=1~\mu$m). Initial conditions are those of rest at the origin of coordinates.}
\label{fig7}
\end{figure}

\subsection{Acceleration from rest}\label{sec:rest}

In laser-assisted atomic ionization, an electron may be produced in a state of near rest (Hu \& Starace, 2002; Maltsef \& Ditmire, 2003). When subsequently subjected to the fields of the chirped pulse, such an electron may be accelerated in vacuum. If the front of the pulse reaches the point at which the electron is {\it born} at the same time, its subsequent dynamics will be dominated by the fields of such a pulse. This is clearly an idealized state of affairs and a probably difficult scenario to realize experimentally. Nevertheless, several aspects of the electron motion will be discussed here based on the solutions to the relativistic equations of motion.

Note first the synchronized rise in the kinetic energy of the electron in Fig. \ref{fig7}, during interaction with the part of the pulse that has been distorted by the chirp. That {\it accelerating} part of the pulse witnesses a substantial decrease in the frequency due to the negative chirp. The electron interacts with an essentially quasi-static electric field and gains about 4.53841 GeV from it, for the parameter set used and $b=-0.01$. Interaction with the higher frequency parts of the pulse results in no further energy gain or loss.

\begin{figure}
\includegraphics[width=8cm]{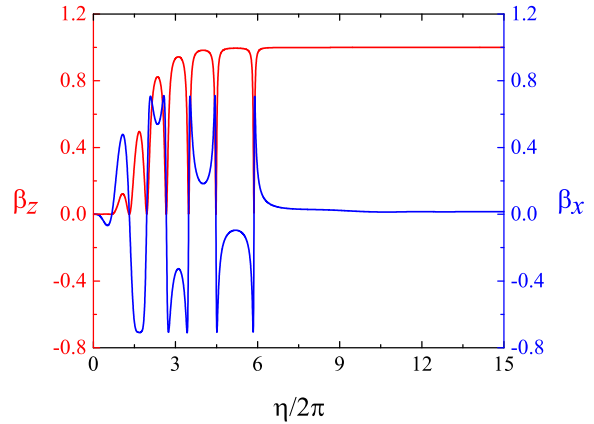}
\caption{(Color online) Evolution in $\eta$ of the components, $\beta_z$ and $\beta_x$, of the velocity vector, normalized by the speed of light in vacuum, of the electron whose kinetic energy is shown in Fig. \ref{fig7}. }
\label{fig8}
\end{figure}

Figure \ref{fig8} shows the $x-$ and $z-$components of the velocity of the electron, scaled by $c$, while it is interacting with the pulse. In the absence of a force $y-$component, the electron motion is confined to the $xz-$plane and, thus, a $\beta_y$ is absent. Note that during the initial stage of interaction with the pulse, $\beta_x$ only rises, due to the $x-$ component of the force exerted by the laser's oscillatory electric field. Soon after that, the $z-$component of the $\bm{v}\times\bm{B}$ force of the laser begins to impart axial kicks to the electron. Thus, while $\beta_x$ continues to exhibit essentially equal-amplitude oscillations for a while, $\beta_z$ oscillates between zero and an increasing positive value which approaches unity quickly as the electron is left behind the pulse. Note that the exit value of $\beta_x$ is nonzero, which shows that the electron will suffer net deflection from pure axial motion (ponderomotive scattering). A scattering angle, indicating the direction in which the electron will emerge after interaction with the pulse has ceased, may be defined by

\begin{equation}\label{theta}
  \theta_{sc}=\arctan\left[\frac{\beta_{x, exit}}{\beta_{z, exit}}\right],
\end{equation}
where $\beta_{x, exit}=\beta_x(\eta_f)$ and $\beta_{z, exit}=\beta_z(\eta_f)$. Using the exit values $\beta_x(\eta_f)\sim0.0150046$ and $\beta_z(\eta_f)\sim0.999887$ results in a scattering angle of $\theta_{sc}\sim0.859732^\circ$.

\begin{figure}
\includegraphics[width=8cm]{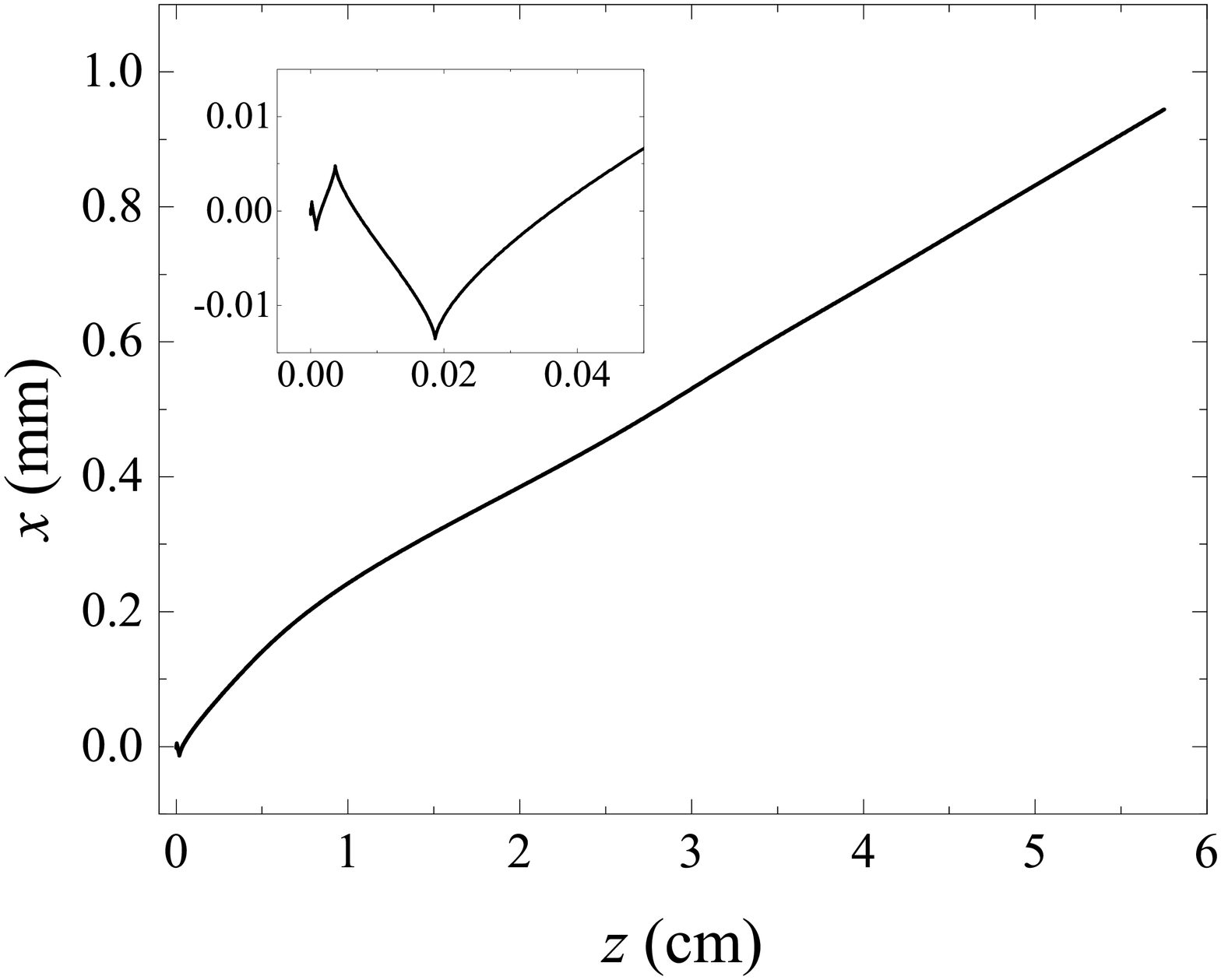}
\caption{2D trajectory of the single electron whose kinetic energy is shown in Fig. \ref{fig7}. The inset is a zoom-in on the part of the trajectory which results from interaction with the first few laser cycles.}
\label{fig9}
\end{figure}

As has just been pointed out, the electron trajectory, while interacting with the laser pulse, is two-dimensional (in the $xz-$plane). A sample trajectory is shown in Fig. \ref{fig9}. Initial interaction with the linearly-polarized pulse results in the familiar structure shown in the inset. After these initial oscillations the electron follows an essentially straight-line trajectory while interacting with the quasi-static part of the electric field. For the parameter set employed, the calculation returns exit transverse and axial excursions $\Delta x\sim 0.944792$ mm, and $\Delta z\sim 5.75264$ cm, respectively.

Performance of a conventional accelerator is often indicated by quoting its {\it average acceleration gradient}. An estimate of this quantity may be obtained by dividing the exit kinetic energy (see Fig. \ref{fig7}) by the exit axial excursion (see Fig. \ref{fig9})
\begin{equation}\label{grad}
  \bar{G} \equiv \frac{\Delta K}{\Delta z},
\end{equation}
where $\Delta K=K(\eta_f)-K(\eta_0)$ and $\Delta z=z(\eta_f)-z(\eta_0)$. For example, a natural limit of $\sim 100$ MeV/m exists on the average gradient of a conventional electron linear accelerator. For the electron accelerated from rest by the present scheme, our results show that $\bar{G}\sim78.8926$ GeV/m, or close to three orders of magnitude times the conventional limit.

\subsection{Acceleration from axial injection}\label{sec:axial2}

\begin{figure}
\includegraphics[width=8cm]{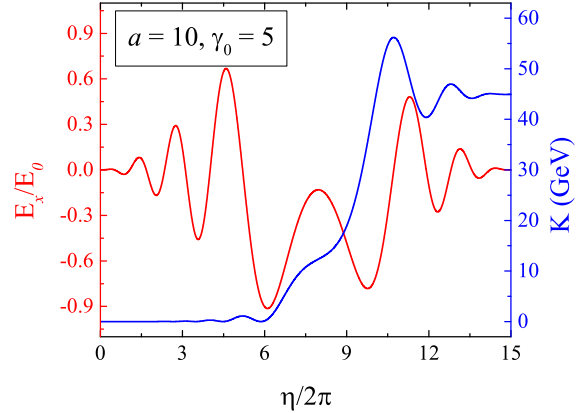}
\caption{(Color online) Same as Fig. \ref{fig7}, but for an electron injected axially with an initial kinetic energy $K_0\sim2.044$ MeV ($\gamma_0=5$).}
\label{fig10}
\end{figure}

An already fast electron, pre-accelerated by a table-top linear accelerator or an electron gun, may be subjected to a static magnetic field in order to bend its trajectory prior to shining the laser pulse on it. This can approximate the case of axial injection, to be discussed here. Again, the assumption is made that the front end of the pulse will catch up with the electron at $t=0$, exactly at the origin of the coordinate system. Following the pattern of the previous subsection, the normalized electric field of the high-intensity chirped pulse is shown in Fig. \ref{fig10}, together with the evolving kinetic energy of the electron during interaction. In addition to exhibiting the same general features as in Fig. \ref{fig7}, Fig. \ref{fig10} shows an increase in the electron's kinetic energy from $K_0\sim2.044$ MeV ($\gamma_0=5$) to $K_{exit}\sim44.9277$ GeV, or more than four orders of magnitude.

\begin{figure}[t]
\includegraphics[width=8cm]{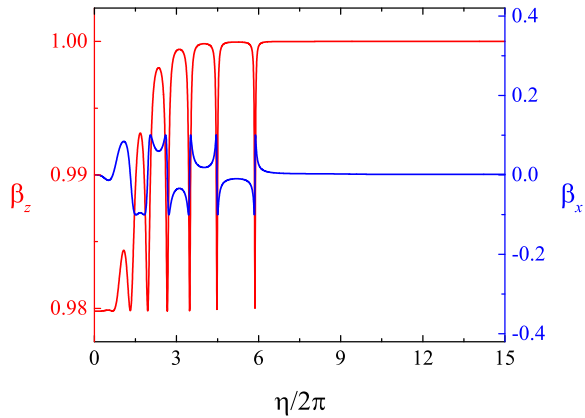}
\caption{(Color online) Same as Fig. \ref{fig8}, but for axial injection with $\gamma_0=5$.}
\label{fig11}
\end{figure}

Evolution of the electron's velocity during interaction with the chirped pulse is shown in Fig. \ref{fig11}. Here, too, subsequent motion is two-dimensional (in the polarization plane of the laser). As for the case of acceleration from rest, $\beta_x$ starts from zero due to the strong action of the oscillating $E_x$ of the laser pulse on it and oscillates between roughly --0.1 and 0.1. Upon exit from the interaction region $\beta_x\sim0.00151586$. On the other hand, $\beta_z$ starts from a relativistic value, oscillates for some time between that value and an increasing maximum, due to the combined action of $E_z$ and $(\bm{v}\times\bm{B})_z$, and finally approaches unity as a result of interaction with the quasi-static electric field. It is eventually left behind the pulse moving at $\beta_z\sim0.999999$.

\begin{figure}[t]
\includegraphics[width=8cm]{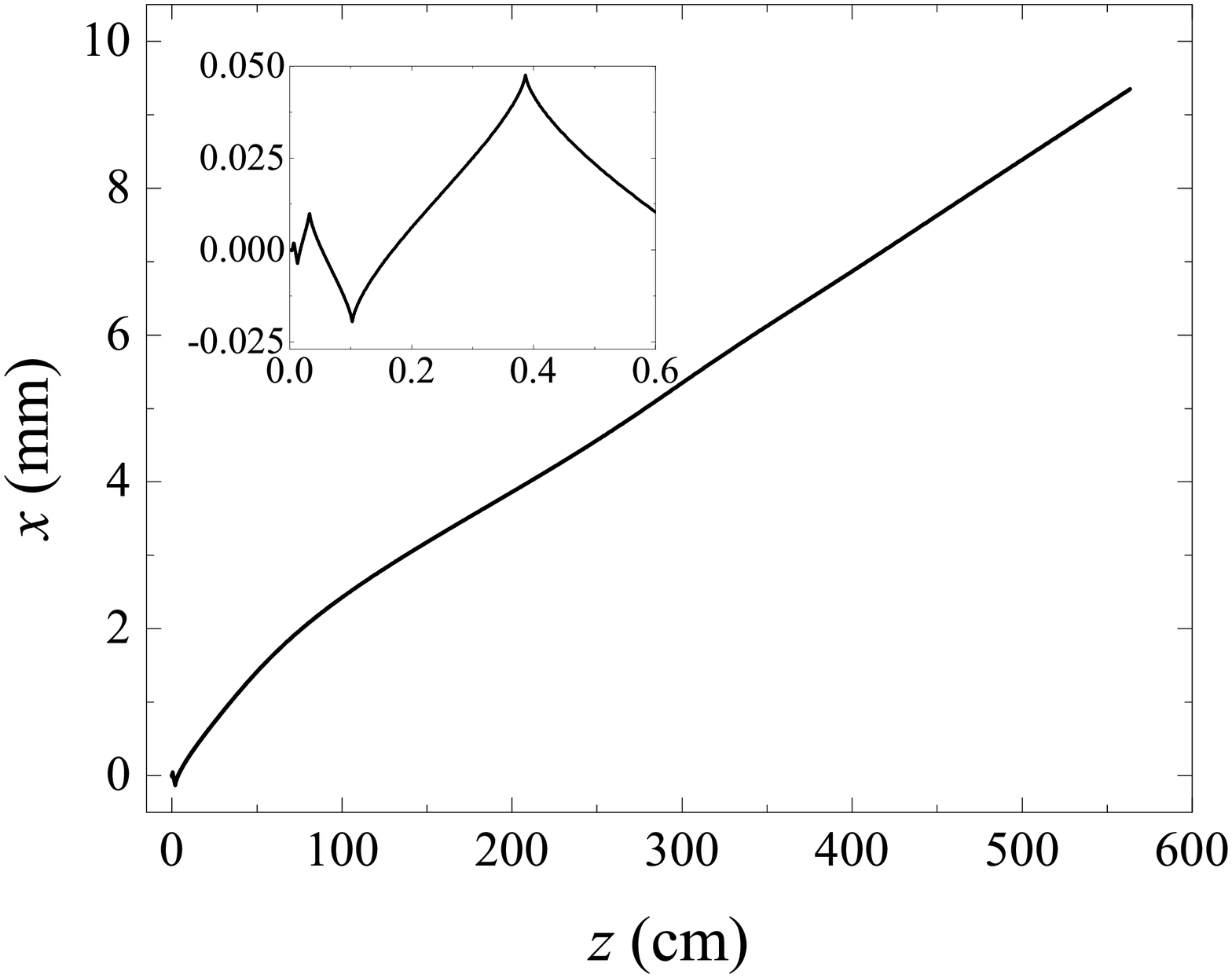}
\caption{Same as Fig. \ref{fig9}, but for axial injection with $\gamma_0=5$.}
\label{fig12}
\end{figure}

Finally, the 2D trajectory for the case of axial injection is shown in Fig. \ref{fig12}. The trajectory looks similar in overall features to the corresponding one displayed in Fig. \ref{fig9} for the case of acceleration from rest. The dimensions in this case, however, are roughly ten-fold transversely and a hundred-fold axially, as a result of the initial forward momentum. Note that the electron undergoes the transverse and axial displacements $\Delta x\sim9.35248$ mm and $\Delta z\sim5.63773$ m, respectively. The latter, when combined with $K_{exit}\sim44.9277$ GeV, yields $\bar{G}\sim 7.96911$ GeV/m. Note further that, although the case of axial injection results in roughly ten times overall exit kinetic energy for the electron, compared with the case of initial rest at the origin, the acceleration gradient is almost ten times less in the latter case compared with the former. The initial forward momentum results in a much larger axial excursion, in this case, than in the case of acceleration from rest.

\subsection{Acceleration from sideways injection}\label{sec:sideways2}

It seems much easier, from an experimental perspective, for the electron to be injected sideways (at an angle $\zeta_0$ with respect to the direction of pulse propagation). This configuration will be discussed next. The injected electron has initial transverse and axial momenta. Thus, its subsequent motion will not be confined to the $xz-$plane. In fact, the trajectory will be 3D, in spite of the fact that the initial injection is made within the $xz-$plane. This is due to the influence of the magnetic field of the laser pulse. Discussions will now be presented of the case of injection at $\gamma_0=5$, but at the angle $\zeta_0=5^\circ$.

\begin{figure}[t]
\includegraphics[width=8cm]{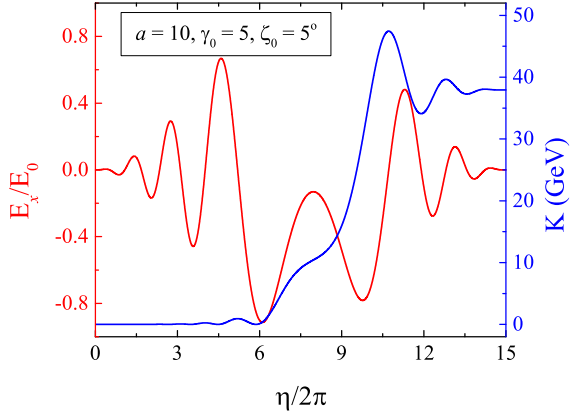}
\caption{(Color online) Same as Fig. \ref{fig7}, but for an electron injected sideways at the angle $\zeta_0=5^\circ$ with the direction of pulse propagation, and an initial kinetic energy $K_0\sim2.044$ MeV ($\gamma_0=5$).}
\label{fig13}
\end{figure}

The normalized electric field of the pulse and evolution of the kinetic energy of the electron are shown in Fig. \ref{fig13}. The fact that the electron, in this case, is injected with less initial axial momentum, than in the case of purely axial injection when the same parameters are used, will result in lower exit kinetic energy for the accelerated electron; see Eq. (\ref{gamma}). The exit kinetic energy that may be read off of Fig. \ref{fig13} is about 38 GeV, compared to $\sim45$ GeV for an axially injected electron.

\begin{figure}[b]
\includegraphics[width=8cm]{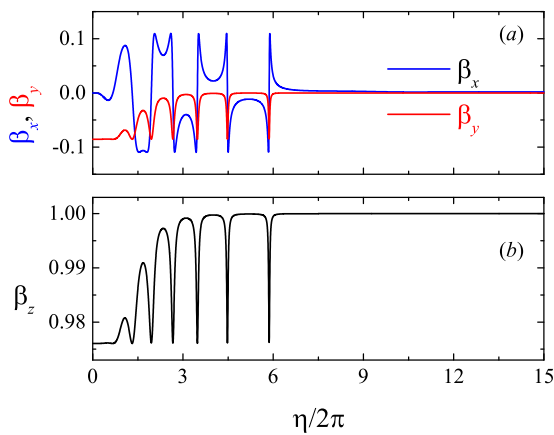}
\caption{(Color online) Same as Fig. \ref{fig11}, but for an electron injected sideways at the angle $\zeta_0=5^\circ$ with the direction of pulse propagation, and an initial kinetic energy $K_0\sim2.044$ MeV ($\gamma_0=5$).}
\label{fig14}
\end{figure}

Components of the scaled velocity vector $\bm{\beta} $ are shown in Fig. \ref{fig14}. As expected, $\beta_x$ starts initially from a value of zero, oscillates during interaction and retains a small exit value, $\sim0.00179557$ for the parameter set employed. On the other hand, the $y-$ and $z-$components, $\beta_y$ and $\beta_z$, have nonzero initial values, which help them accumulate successive increases after interaction with every laser cycle. Partly due to the fact that the injection angle is too small and, hence, the initial value of $\beta_z$ is greater than that of $\beta_y$, the exit values of these two components are widely different ($\beta_y\to-5.75237\times10^{-6}$ and $\beta_z\to0.999998$). Using the exit values of $\beta_x$ and $\beta_z$ in Eq. (\ref{theta}) gives $\theta_{sc}\sim0.102879^\circ$. In conclusion, the electron emerges from interaction with the pulse essentially axially.

\begin{figure}[t]
\includegraphics[width=8cm]{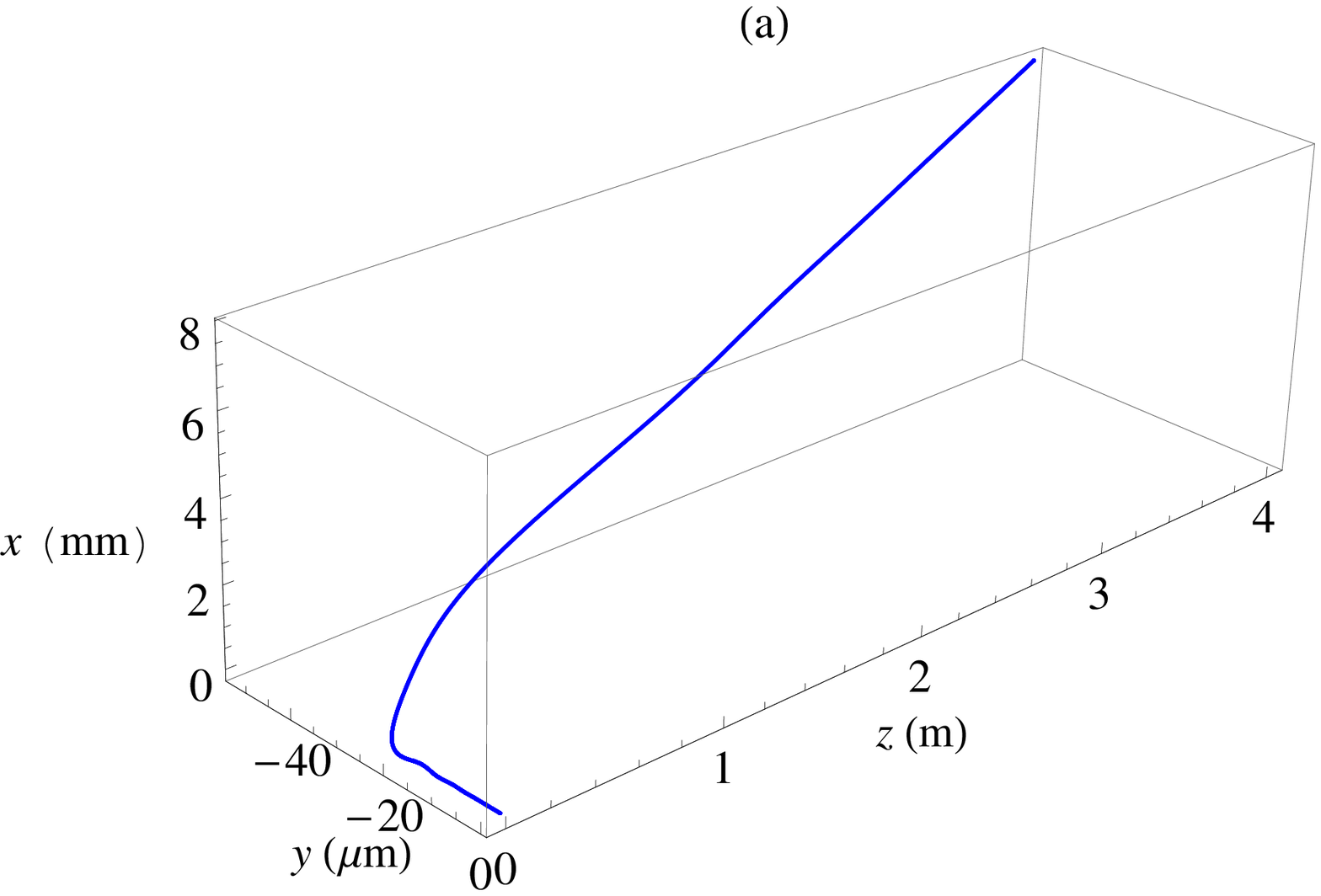}
\includegraphics[width=8cm]{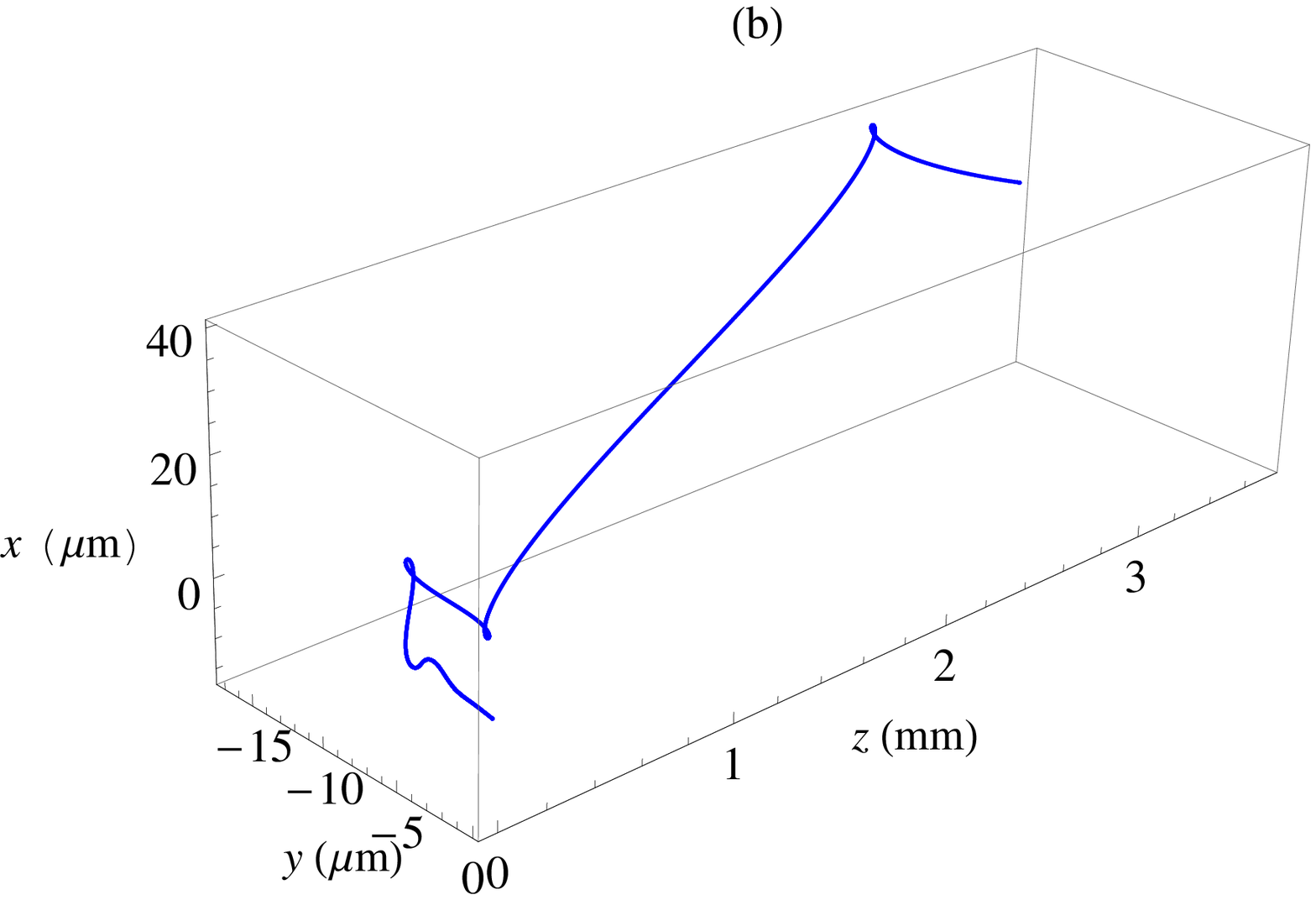}
\caption{(Color online) 3D trajectory of a single electron submitted to a chirped laser pulse ($b=-0.01$). In (b) we zoom-in on part of the trajectory followed during roughly the first one-third of the interaction time with the pulse. Injection is at $\zeta_0=5^\circ$ and $\gamma_0=5$ ($K_0\sim2.044$ MeV). The laser parameters are: $a=10, \lambda_0=1~\mu$m, and $\tau=50$ fs.}
\label{fig15}
\end{figure}

Sideways injection of the electron leads to the 3D trajectory shown in Fig. \ref{fig15}. Details of the part of that trajectory which result from interaction with the first few laser cycles are displayed in Fig. \ref{fig15}(b). The existence of turning points in the $x-$ and $y-$coordinates is quite evident. More clearly pronounced is the forward drift in $z$. The following exit coordinates have been calculated: $x_{exit}\sim7.89547$ mm, $y_{exit}\sim-53.5223~\mu$m, and $z_{exit}\sim4.01806$ m. Therefore, using the calculated exit kinetic energy of $K_{exit}\sim37.9288$ GeV, Eq. (\ref{grad}) yields $\bar{G}\sim9.43905$ GeV/m, or about two orders of magnitude better than the limit of a conventional linear accelerator.

\section{Discussion and conclusions}\label{sec:discussion}

Dynamics of a single electron submitted to a high-intensity chirped laser pulse has been the subject of investigation in this work. The aim has been mainly pedagogical: to gain more insight into the process of electron acceleration by a chirped laser pulse. A simple plane-wave model representation of the electromagnetic fields of the pulse has been employed. Three different initial conditions on the position and velocity of the electron have been considered: rest at the origin, injection axially and injection at an angle $\zeta_0$ relative to the direction of pulse propagation. Working equations of general applicability have been derived and used directly to study some aspects of the electron dynamics, or else have served as a basis for benchmarking the numerical simulations that were conducted to investigate further more sophisticated aspects.

The case of axial injection has been taken up first, using parameters the same as ones employed in a similar study, albeit for a non-chirped laser pulse (Hartemann {\it et al.}, 1995). The aim has been to show that our chirped-pulse-based investigations would produce the same results in the appropriate limit, both analytically and numerically. This aim has been fully accomplished.

Then more thorough investigations were conducted, which typically started by scanning a small portion of the chirp parameter space for values that would lead to sizable energy gains by the electron from interaction with the pulse. It has been shown that the chirp parameter values which lead to maximum gain are not affected by the change of conditions on the electron initial injection. The same values of $b$ maximize the gain in all considered configurations.

In an actual experiment, one would be interested in a number of important details, such as the actual trajectory, the ejection dynamics and, most importantly, the exit kinetic energy of the electron. In our single-particle calculations, synchronous motion of the electron, which is believed to be responsible for the energy gain, has been demonstrated by showing evolution of its kinetic energy in a typical event together with the normalized electric field seen by it. It has been shown, in all cases considered, that substantial energy gain takes place always during interaction with a quasi-static part of the field generated by chirping the pulse frequency appropriately.

Evolution of the velocity components during interaction have been studied. More importantly, the exit electron velocities were found in each example and a scattering angle has been calculated. In all cases, the particles were found to be ejected within a small cone about the propagation direction of the pulse. In other words, while this demonstrates ponderomotive scattering, it also predicts that a bunch or beam of electrons would suffer little spatial diffraction.

To gain further insight into the particle-field interactions, actual trajectories for the various investigated events have been calculated. Straightforward arguments, based on the general equations of motion, led to the conclusion that the trajectories in the cases of acceleration from rest at the origin or from axial injection would be 2D, motion of the electron would be confined to the polarization plane of the pulse. Sideways injection, however, has been shown to lead to 3D trajectories.

Finally, it should be borne in mind that the model adopted in this work is plane-wave in character. As such, some of the results may turn out to be overly exaggerated. The kinds of laser peak intensities used in our calculations can only be reached by tightly focused pulses. As is well known, a tightly focused gaussian beam, for example, is plane-wave in character only on the focal plane or else too far away from such a plane. Thus, the energy gains of several tens of GeV and the acceleration gradients of a few GeV/m, predicted in this work, may be overestimates of what is expected to emerge from a more realistic model, like that of a tightly focused gaussian pulse.\\

\flushleft{\bf ACKNOWLEDGMENTS}\\

\flushleft{YIS acknowledges support for this work from an American University of Sharjah Faculty Research Grant (FRG-III).}\\

\flushleft{\bf REFERENCES}\\
~~~\\

\textsc{Esarey, E., Sprangle, P., \& Krall, J.} (1995). Laser acceleration of electrons in vacuum. {\it Phys. Rev. E} {\bf 52}, 5443.

\textsc{Galow, B., Salamin, Y., Liseykina, T., Harman, Z., \& Keitel, C.} (2011). Dense Monoenergetic Proton Beams from Chirped Laser-Plasma Interaction. {\it Phys. Rev. E} {\bf 107}, 185002.
 
\textsc{Gupta, D. N. \& Suk, H.} (2007). Electron acceleration to high energy by using two chirped lasers. {\it Laser Part. Beams} {\bf 25}, 31.

\textsc{Hartemann, F., Fochs, S., Sage, G. L., Luhmann, N., Woodworth, J., Perry, M., Chen, Y. J. \& Kerman, A. K.} (1995). Nonlinear ponderomotive scattering of relativistic electrons by an intense laser field at focus. {\it Phys. Rev. E} {\bf 51} 4833.

\textsc{Hu, S. X. \& Starace, A. F.} (2002). GeV Electrons from Ultraintense Laser Interaction with
	Highly Charged Ions. {\it Phys. Rev. Lett.} {\bf 88}, 245003.\\

\textsc{Kawata, S., Kong, Q., Miyazaki, S., Miyauchi, K., Sonobe, R., Ssksi, K., Nakajima, K., Masuda, S., Ho, Y. K., Miyanaga, N., Limpouch, J. \& Andreev, A. A.} (2005). Electron bunch acceleration and trapping by ponderomotive force of an intense short-pulse laser. {\it Laser Part. Beams} {\bf 23}, 61.

\textsc{Lawson, J. D.} (1979). Lasers and accelerators. 
	{\it IEEE Trans. Nucl. Sci.} {\bf NS-26}, 4217.\\

\textsc{Malka, V., Lefebvre, E. \& Miquel, J. L.} (1997). Experimental observation of electrons accelerated in vacuum to relativistic energies by a high-intensity laser. {\it Phys. Rev. Lett.} {\bf 78}, 3314.\\
 
\textsc{Maltsev, A. \& Ditmire, T.} (2003). Above Threshold Ionization in Tightly Focused, Strongly 	Relativistic Laser Fields. {\it Phys. Rev. Lett.} {\bf 90}, 053002.\\

\textsc{Pang, J., Ho, Y. K., Yuan, X. Q., Cao, N., Kong, Q., Wang, P. X., Shao, L., Esarey, E. H. \& Sessler, A. M.} (2002). Subluminous phase velocity of a focused laser beam and vacuum laser acceleration. {\it Phys. Rev. E} {\bf 66}, 066501.\\

\textsc{Plettner, T., Byer, R. L., Colby, E. , Cowan, B., Sears, C. M. S., Spencer, J. E.,
    \& Siemann, R. H.} (2005). Visible-laser acceleration of relativistic electrons in a 
		semi-infinite vacuum. {\it Phys. Rev. Lett.} {\bf 95}, 134801.\\
 
\textsc{Salamin, Y. I.} (2006). Electron acceleration from rest in vacuum by an axicon Gaussian laser beam. {\it Phys. Rev. A} {\bf 73}, 043402.\\

\textsc{Salamin, Y. I.} (2012). Net electron energy gain from interaction with a 
	chirped “plane-wave” laser pulse. {\it Phys. Lett. A} {\bf 376}, 2442.\\

\textsc{Salamin, Y. I., Faisal, F. H. M. \& Keitel, C. H.} (2000). Exact analysis of ultrahigh laser-induced acceleration of electrons by cyclotron autoresonance. {\it Phys. Rev. A} {\bf 62}, 053809.\\

\textsc{Salamin, Y. I., Harman, Z. \& Keitel, C. H.} (2008). Direct high-power laser acceleration of ions for medical applications. {\it Phys. Rev. Lett.} {\bf 100}, 155004.\\
 
\textsc{Salamin, Y. I. \& Keitel, C. H.} (2002). Electron acceleration by a tightly focused laser beam. {\it Phys. Rev. Lett.} {\bf 88}, 095005.\\
 
\textsc{Scully, M. O. \& Zubairy, M. S.} (1991). Simple laser accelerator: Optics and particle dynamics. {\it Phys. Rev. A} {\bf 44}, 2656.\\

\textsc{Shao, L., Cline, D., Ding, X., Ho, Y. K., Kong, Q., Xu, J. J., Pogorelsky, I., Yakimenko, V., \&Kusche, K.} (2013). Simulation prediction and experiment setup of vacuum laser acceleration at Brookhaven National Lab-Accelerator Test Facility. 
{\it Nucl. Inst. \& Meth. Phys. Res. A} {\bf 701}, 25. \\
  
\textsc{Singh, K. P.} (2005). Electron acceleration by a chirped short intense 
	laser pulse in vacuum. {\it Appl. Phys. Lett.} {\bf 87}, 254102.\\
 
\textsc{Sohbatzadeh, F. \& Aku, H.} (2011). Polarization effect of a chirped Gaussian 
	laser pulse on	the electron bunch acceleration. {\it J. Plasma Phys.} {\bf 77}, 39.\\
 
\textsc{Sohbatzadeh, F., Mirzanejhad, S. \& Aku, H.} (2009). 
	Synchronization scheme in electron vacuum acceleration by a chirped 
	Gaussian laser pulse. {\it Phys. Plasma} {\bf 16}, 023106.\\
 
\textsc{Sohbatzadeh, F., Mirzanejhad, S., Aku, H. \& Ashouri, S.} (2010). 
	Chirped Gaussian laser beam parameters in paraxial approximation. 
	{\it Phys. Plasma} {\bf 17}, 083108.\\
 
\textsc{Sohbatzadeh, F., Mirzanejhad, S. \& M. Ghasemi} (2006). Electron acceleration 
	by a chirped 	Gaussian laser pulse in vacuum. {\it Phys. Plasma} {\bf 13}, 123108.\\
 
\textsc{Umstadter, D., Kim, J., Esarey, E., Dodd, E. \& Neubert, T.} (1995). Resonantly laser-driven plasma waves for electron acceleration. {\it Phys. Rev. E} {\bf 51}, 3484.\\

\textsc{Wang, P. X., Ho, Y. K., Yuan, X. Q., Kong, Q., Cao, N., Sessler, A. M., Esarey, E., Nishida, Y.} (2001). Vacuum electron acceleration by an intense laser. {\it Appl. Phys. Lett.} {\bf 78}, 2253.\\

\textsc{Woodward, P. M.} (1947). A method of calculationg the field over a plane. 
	{\it J. Inst. Elect. Eng.} {\bf 93}, 1554.\\
 
\textsc{Yanovsky, V., Chvykov, V., Kalinchenko, G., Rousseau, P., Planchon, T., Matsuoka, T., Maksimchuk, A., Nees, J., Cheriaux, G., Mourou, G. \& Krushelnick, K.} (2008). Ultra-high intensity- 300-TW laser at 0.1 Hz repetition rate. {\it Opt. Expess} {\bf 16}, 2109.\\

\textsc{Xie, Y. J., Wang, W., Zheng, L., Zhang, X. P., Kong, Q., Ho, Y. K. \& Wang, P. X.} (2010). Field structure and electron acceleration in a slit laser beam. {\it Laser Part. Beams} {\bf 28}, 21.\\

\textsc{Xu, J. J., Kong, Q., Chen, Z., Wang, P.X., Wang, W., Lin, D. \& Ho, Y. K.} (2007). Polarization effect of fields on vacuum laser acceleration. {\it Laser Part. Beams} {\bf 25}, 253. 
\end{document}